\documentstyle[prl,preprint,aps]{revtex}

\begin{document}
\draft
\preprint{\it Submitted to Physics Letters A}
\title
{Magnetic softening of Young's modulus of amorphous Fe$_{90}$Zr$_{10}$. \\}

\author{S. H. Park, Y. H. Jeong\cite{correspondence} \\
{\it Department of Physics, Pohang University of Science and Technology \\
 Pohang, 790-784, S. Korea \\} }
\author{K. Nahm, C. K. Kim \\
{\it Department of Physics, Yonsei University, Seoul, 120-749, S. Korea} }
\date{\today}
\maketitle

\begin{abstract}
The Young's modulus and the internal friction of amorphous
Fe$_{90}$Zr$_{10}$ alloy were measured near the Curie
temperature  using the vibrating reed technique.
The modulus shows softening around $T_c\approx 227K$ and the
internal friction  undergoes drastic increase at $T_c$.
It is found that both the Young's modulus and the reciprocal of
internal friction are inversely proportional to the magnetic
susceptibility in the paramagnetic phase.
\end{abstract}
\pacs{PACS numbers: 75.50.K, 62.20.D, 75.80}

\bibliographystyle{prl}

Amorphous iron-rich Fe-Zr alloys near 90 at. \% Fe have received
considerable attention recently because of their unusual magnetic
properties. These alloys have been extensively
studied by various methods such as magnetic,\cite{Saita86,Coey87,Heller86}
thermal,\cite{Coey87,Obi82} transport,\cite{Obi82,Fukamichi82,Dahlberg84}
M\"ossbauer,\cite{Heller86,Ryan87} and neutron scattering\cite{Rhyne85}
measurements.  From these measurements,\cite{Coey87} it is known that the
system
undergoes two successive transitions from a paramagnetic to ferromagnetic
at the
Curie temperature T$_c$(with T$_c$ decreasing with increasing iron
concentration above 85 at. \% Fe) and to a spin glass phase
at T$_f$(less than T$_c$).
However, the exact nature of these transitions and magnetic states is still in
controversy.\cite{Saita86,Coey87}

Of particular interest about these alloys is its temperature
dependence of magnetoelastic properties. It was reported that the
a-Fe$_{90}$Zr$_{10}$ shows the {\it Invar} anomaly in the thermal
expansion\cite{Shirakawa}. Despite the fact that  many  magnetic alloys
have been shown to be of Invar type since the first discovery by
Guillaume,\cite{guillaume} the physical  mechanism of the effect has never
been fully
understood and still is the subject of  active research. Although the
name `Invar' originated from an anomaly in the  thermal expansion, Invars are
usually found to have a broad variety of associated anomalies.\cite{Buschow}  A
large magnetic softening in the elastic modulus,  for example, frequently
accompanies the thermal expansion anomaly near the Curie
temperature.\cite{Wohlfarth76}  Since most of the works on the magnetoelastic
properties of Invar alloys have been done on crystalline
alloys,\cite{Buschow,example} it
is of value to study the elastic properties of  a-Fe$_{90}$Zr$_{10}$ which is
an amorphous invar.  This in turn would shed light on the magnetic nature of
the
system. In this paper, we report the results of our investigation on the
variation
of the Young's modulus of  a-Fe$_{90}$Zr$_{10}$ as a function of temperature.

A frequently used method to measure the elastic modulus of solids is
ultrasonic measurements. However, this method is not suitable for
a-Fe$_{90}$Zr$_{10}$ due to the fact that samples are of ribbon type as is
usual
with metallic glasses. Instead the vibrating reed
technique,\cite{Golding,Gupta} which is an excellent method for the
measurements
of the elastic modulus of metallic films, was used to investigate the
temperature
dependence of the  Young's modulus and the internal friction of amorphous
Fe$_{90}$Zr$_{10}$  ribbons from 100 K up to 300 K. Since the details of the
experiment will be reported elsewhere,\cite{Bae} we only briefly describe the
sample preparation method and the vibrating technique here.

The Fe-Zr alloy ingot 10 at \% Zr was prepared by the arc melting
in a purified argon atmosphere.
The ribbon, about 2 mm wide and 20 $\mu$m thick, was prepared by
the rotating drum technique under an argon atmosphere. Small parts of the
amorphous ribbon were investigated by x-ray diffraction and found
to have x-ray diagrams characteristic of the amorphous state, where
sharp diffraction lines indicating the presence of the crystalline phases
were absent. No deviation from the nominal concentration within the
uncertainty of the measurement of 1 at \% could be detected with
microprobe analysis.

In measuring the Young modulus, one end of the sample was clamped between two
copper flats and the  free end was driven electrostatically by a sinusoidal
voltage
of frequency $f/2$. Since in an electrostatic drive without bias
voltage the resulting force is proportional to the square of the
voltage, oscillation occurs at frequency $f$.
When $f$ matches with one of the resonant frequencies $f_n$ of the
ribbon, then resonance occurs. For flexural modes the resonant frequencies
are given
by\cite{flexural}
\begin{eqnarray}
f_n=\frac{d}{4\pi \sqrt{3}}\left (\frac{\beta_n}{L}\right ) ^2\sqrt
{\frac{E}{\rho}}
\end{eqnarray}
where $d$ is the thickness of the ribbon, $L$ its length, $E$ the Young's
modulus, and $\rho$ the mass density. $\beta_n$'s are
constants which satisfy the equation
$\cos \beta_n \cosh \beta_n +1=0$ and the values of the first few ones are
1.875, 4.694, 7.855. The amplitude of the vibration was detected capacitively.
The setup of the sample cell is shown in the inset of Fig.~\ref{fig:mod}.
As the reed
vibrates, the reed and the detecting electrode forms a time varying
capacitance whose ac
amplitude is  proportional to the displacement of the reed. This variation
of the
capacitance was picked up with a lock-in amplifier.
The resonant frequency and the internal friction $Q^{-1}$ were
obtained from the shape of the resonance curve by sweeping the driving
frequency
through the resonance and fitting the data with the damped
harmonic oscillator form. During each frequency sweep which took about 5
minutes
temperature was controlled with stability of 5 mK.

In order to locate the transition point we first carried out the measurements
of
the magnetic susceptibility using the mutual inductance ac-method.\cite{Moon}
An oscillating magnetic field of less than 5 mOe at 48 Hz was applied along the
length of a 1 cm long sample to minimize demagnetizing effects. Temperature was
stable at a given set point within 5 mK. Fig.~\ref{fig:sus} is the plot of the
susceptibility against temperature. As temperature falls toward $T_c$,
$\chi$ rises
sharply and goes through a maximum called  a Hopkinson peak.  Fitting the
data of the
paramagnetic phase with the usual power law, $\chi\,\sim
|T\,-\,T_c|^{-\gamma}$, yields
T$_c$ = 227 K and $\gamma$ = 1.39. These values are in good accordance with
previous
measurements.\cite{Kaul86}  Notice that the data deviate from the fit in
the vicinity of
$T_c$. This is due to the unavoidable demagnetizing effect and one should
regard the
solid line as  representing the intrinsic values.

In Fig.~\ref{fig:mod} we plot the variation of the
Young's modulus of a-Fe$_{90}$Zr$_{10}$, normalized to that at 300 K, as a
function
of temperature. For this particular sample, the length of the reed was
about 1 cm
which resulted in resonance frequencies $f_1\simeq 106$ Hz and $f_2\simeq
660$ Hz at
300 K. The measurements were done by detecting the fundamental frequency as the
temperature was varied.  The absolute value of the modulus at 300 K was
calculated,
using Eq. (1) to yield $E\simeq 3.5\times 10^{11}$ dyne/cm$^2$ which gives
the sound
velocity $s\simeq 3.3$ km/s.
	The modulus does not vary appreciably as the temperature
is lowered from 300 K to 240 K where the sudden softening starts to set in.
This abrupt softening continues until the temperature reaches T$_c$. In the
ferromagnetic phase the modulus still decreases, with the smaller slope, as
the temperature is reduced further. The behavior of the Young's modulus in the
ferromagnetic phase represents the interesting physics; however, we defer the
discussion of this point to the full paper\cite{Bae} and restrict our
discussion on
the paramagnetic phase in this letter.

It is noted that the modulus softening occurs mostly above T$_c$,
i.e., in the paramagnetic phase. This behavior cannot be explained in terms
of the Landau theory which is frequently invoked to account for magnetic
softening of elastic constants.\cite{Landau} Rather, this result strongly
indicates
the importance of coupling between the spin fluctuations and phonons. Thus,
it would
be appropreate to investigate the relationship between the elastic modulus
and the
magnetic susceptibility which is a measure of the spin fluctuations.
Fig.~\ref{fig:mod-sus} is the plot of the Young's modulus against the inverse
of
susceptibility. Self-evident is the linear relationship between E and
1/$\chi$ in the
temperature range near T$_c$ where most of the softening of E and the
increase in $\chi$
occur. This is, to our knowledge, the first time this feature is
displayed in metallic magnetic systems.

Since the modulus is strongly affected by the magnetic phenomenon occuring
in the
system, it is reasonable to expect that the internal
friction would also reflect the magnetic ordering. To see this effect,
however, one
must take into account the thermoelastic relaxation\cite{Zener}, since it
usually
is an important loss mechanism in the vibrating reed.
According to Golding\cite{Golding}, damping
caused by the thermoelastic relaxation is given by
\begin{eqnarray}
Q^{-1}_{th} &=&TE_T\alpha _p^2/C_p\frac{\omega\tau}{1+\omega^2\tau^2}
\label{eq:Q}
\end{eqnarray}
with $\tau = (d/\pi)^2C_p/\kappa\,$ where $d$ is the thickness of the sample.
Here $E_T, \alpha_p, \kappa$ and $C_p$ are the isothermal Young's modulus,
linear thermal expansivity, thermal conductivity, and isobaric specific
heat respectively.
Evaluating Eq.~(\ref{eq:Q}) by putting the values of relevant parameters,
we obtain
$Q^{-1}_{th}\,\sim 6.5\times 10^{-8}\,.$ Since this value is very much
smaller than
those measured experimentally, we can safely disregard thermoelastic
relaxation effects.
Note that the facts that a-FeZr is an Invar, i.e.,
$\alpha$ is small, and has  metallic thermal conductivity allow the
vibrating reed
technique to be an ideal method to probe any change in the internal
friction due to the
intrinsic effect associated with the magnetic ordering. This is to be
contrasted to the
situation with EuO where the thermoelastic relaxation is
dominant.\cite{Golding}

The inset of Fig.~\ref{fig:Q-sus} displays the internal friction as a
function of
temprature. Indeed there is a steep increase in the internal friction $Q^{-1}$
in conjunction with the softening of the modulus near T$_c$.  Remembering
that the steep
increases are present both in the susceptibility and in the internal
friction near
T$_c$, we have plotted in Fig.~\ref{fig:Q-sus} the inverse of the internal
friction,
$1/Q^{-1}$, versus the reciprocal of the susceptibility.  Here again the linear
relationship is obeyed in the same temperature range as was with the
Young's modulus and
the susceptibility.

At present we are not aware of any theory  which is fully capable of
explaining our
experimental results. However, the theory due to Kim\cite{Kim} which
stresses the
role of the electron-phonon interaction in metallic magnetism seems to be
relevant.
According to this theory, the dynamic screening of phonons
by the mutually interacting electrons is the main cause of all the
anomalies of metallic
magnetism; assuming the jellium model for a metal\cite{Pines} and using the
mean field
approximation, he showed how the phonon frequency is influenced by the
electronic
behaviors. The relationship between the velocity and the  attenuation of
sound and the
magnetic susceptibility is especially simple in the paramagnetic phase.
That is,
the sound velocity s  and the sound attenuation $\Gamma$ are given by
\begin{eqnarray}
s/s_0&=&[\xi+\chi_0/\chi]^{1/2} \nonumber \\  \Gamma/\Gamma_0&=&(s_0/s)^2
\label{eq:sound}
\end{eqnarray}
where
$\chi$ is the Stoner susceptibility and $s_0$ is the Bohm-Staver sound
velocity of the
system. $\chi_0, \Gamma_0$ and $\xi$ are constants characterizing the
system and of
these constants $\xi$ is the only parameter which can be temperature dependent.
$\xi$
quantifies  how much the system deviates from the jellium model which
assumes point-like
ionic interactions; in other words, it quantifies short-range steric
interactions
between ions. The wavenumber dependent frequency, $\Omega(q)$, for longitudinal
oscillations of ions is given by
\begin{eqnarray}
\Omega(q)^2\,=\,\Omega _p^2\,+\,\xi s_0^2q^2
\end{eqnarray}
where $\Omega _p$ is the ionic plasma frequency. Thus, one
expects that it would be a very sensitive function of distance between ions and
consequently dominates the temperature dependence of Eq.~(\ref{eq:sound})
in ordinary
situations. However, for Invars $\xi$ may be regarded as a constant, since
$\alpha$ is
extremely small.

Thus, assuming that {\it a}-Fe$_{90}$Zr$_{10}$ is an itinerant
ferromagnetic system,\cite{itinerant}  Kim's theory may appear to correctly
account for
the experimental results since the Young's modulus is proportional to square of
the sound velocity and $\xi$ does not vary much within the narrow
temperature range
near T$_c$ where most of changes in the susceptibility, the sound velocity,
and the
internal friction occur. However, it should be kept in mind that the
susceptibility
appearing in the theory is the Stoner one, which does not have the same
temperature
dependence as the measured one. Therefore, it is premature to conclude that
the theory is
supported by the experimental results. We merely point out that Kim's
theory would be
more suitable for amorphous metals in view of the fact that the theory is
that of free
electrons based on the Jellium model, which was pointed out as a weak point
of the
theory.\cite{critic}

In conclusion, we have discovered that the Young's modulus
and the inverse of the internal friction of a-Fe$_{90}$Zr$_{10}$
alloy are linearly proportional to the reciprocal of the magnetic
susceptibility in the
paramagnetic phase. This behavior seems to suggest the importance of
the electron-phonon interaction in this system. Sytematic study is under way to
see whether this effect is universal in amorphous Invars.

We gratefully acknowledge invaluable discussions with Prof. B. I. Min.
YHJ wish to thank financial supports from Korea Research Foundation and Dong-il
Foundation. This work was partly supported by the BSRI program of POSTECH and
Korea-Russia Collaboration program of KOSEF.

\begin{figure}
\caption{Plot of the susceptibility of a-Fe$_{90}$Zr$_{10}$ against
temperature. The
solid line represents the power law fit to $\chi\,\sim
|T\,-\,T_c|^{-\gamma}$, with
T$_c$ = 227 K and $ \gamma $ = 1.39.} \label{fig:sus}
\end{figure}

\begin{figure}
        \caption{Temperature dependence of the Young's modulus.
        The modulus was normalized to the value (E$_0$) at 300 K. Inset
shows the
capacitive cell used in the present measurements.}
        \label{fig:mod}
\end{figure}

\begin{figure}
        \caption{Plot of Young's modulus versus $1/\chi$. The straight line
is the guide
to the eye.}
        \label{fig:mod-sus}
\end{figure}

\begin{figure}
        \caption{Plot of $1/Q^{-1}$ versus $1/\chi$. Inset displays
$Q^{-1}$ as a
function of temperaature.}
        \label{fig:Q-sus}
\end{figure}


\begin{references}
\bibitem[*]{correspondence}
To whom all the correspondence should be addressed.
\end{references}
\end{document}